# Stability of motion of $N$-body systems

A.A. El-Zant

*Astronomy centre, University of Sussex*

**Abstract.** Complications arising from the non-compact nature of the phase space of $N$-body systems prevent any asymptotic characterization of chaotic behaviour. This leads us to revisit some of the old results concerning geodesic stability on Lagrangian manifolds. These cannot be applied directly but may be useful when properly reinterpreted. A method for doing that is described along with some tests of its validity and possible future applications.

## 1 Introduction

If there's a complete lack of particle-particle correlations, an $N$-body gravitational system is fully describable by a one particle distribution function on the 6 dimensional phase-space. If moreover almost all orbits are regular quasiperiodic then this space divides into a set of 3-D "ergodic components" in the form of KAM tori. A probability density on a 3-torus quickly becomes time independent so that this state of affairs can be fully characterized by stationary solutions of the Collisionless Boltzmann Equation (CBE). If the system is sufficiently far from any chaotic region then neighbouring states cannot be too dissimilar (KAM theory). Thus sufficient near integrability of a $N$-body system is a sufficient condition for both the existence and stability of solutions of the CBE. These properties however are far from understood in the general case.

It can be shown under fairly general assumptions that the dynamics of gravitational $N$-body systems is described by the CBE in the limit of $N \to \infty$ (Braun and Hepp 1977). The question therefore is that of the existence and stability of steady state solutions when a significant amount of chaos is present. Of course chaos in principle does not rule out quasistationary solutions of the CBE. However, chaotic orbits take much longer for their time averaged densities to converge to a steady state value, thus making it much more difficult to construct steady state solutions (if only because they will explore a higher dimensional phase space). They are also much more liable to be affected by discreteness noise, thus although solutions may exist in the infinite $N$ limit they would be unstable to perturbations due to discreteness (ie. for any finite $N$). It is to be noted here that the standard two body relaxation estimates do not guaranty stability against chaotic behaviour since they assume *a priori* that the system being dealt with is integrable (in some sense even linear) and that it remains so under the perturbations due to discreteness. The contradictions arising from this type of analysis have been known for a long time (eg. Kurth 1957) but are still widely ignored.

A characterization of chaos in the $N$-Body problem is an essential first step towards understanding its effect. In section 2 we describe some of the peculiarities of $N$-body systems which demand the use of a local approach, while in section 3 we describe one such method which makes use of local geometric quantities, sections 4 and 5 are devoted to a derivation and practical application of a criterion that appears to be suitable for $N$-body systems.

## 2   What's wrong with this $N$-Body problem?

Central to all attempts at understanding the statistical behaviour of physical systems in terms of microscopic dynamics is the idea of phase space mixing (eg. Sagdeev *et. al.* 1988). If a system is mixing in an area of its phase-space, it will explore that area and reside most of the time in a region corresponding to the largest number of microstates compatible with some macroscopic state. If moreover the phase space is compact and the mixing takes place for all initial conditions then Gibb's ergodic hypothesis is satisfied and a Gibb's measure is defined (Pesin 1989). The system is now in a most probable state. If one is interested in finding out whether such a state is ever reached then it might be useful to use asymptotic quantities to characterize the divergence of trajectories in phase-space (which is the cause of the mixing). One such measure is provided by the Liapunov exponents. A system that has positive exponents for initial conditions in a subset of its phase space has positive Kolmogorov entropy and is therefore mixing and ergodic and approaches statistical equilibrium on this set.

A number of problems arise however if one tries to apply this simple picture to gravitational systems. First, because of the singularities when the distance between any two particles vanishes, phase space cannot be compact. This problem can only be made worse if the system is unbounded as are realistic astronomical systems. Then there's the problem with the long range of the gravitational forces which prevent most probable steady states from being defined in the conventional sense (Padmanabhan 1990). This property is also responsible for collective instabilities that tend to drive gravitational systems towards more asymmetric and clustered states (such as bar instabilities) as opposed to more conventional thermodynamic behaviour where systems evolve towards more symmetric states which equalize their macroscopic parameters (eg. temperature). For closed spherical systems this is illustrated by the well known conclusions of Lynden-Bell and Wood (1968) that for a given volume and mass such configurations evolve towards isothermal states only if the energy is high enough to quench the tendency to cluster caused by the long range forces. Otherwise the system clusters and separates into components with heat effectively flowing from the colder regions to the hotter ones. Such is also believed to be the fate of open systems, except that in this case stars gain enough energy to escape and the system eventually dissolves. A system of escaping point masses (with possibly a few binaries left over) has zero Liapunov exponents, thus strictly speaking a gravitational $N$-body system cannot be chaotic in the asymptotic regime no matter what the transient dynamics had been.

## 3 Towards a local characterization of chaos

The asymptotic characterization of chaos described above has the appealing property of being independent of any structure imposed on the phase space. The existence of the Liapunov exponents is guaranteed by virtue of the Osledec theorem which bypasses the need for a particular metric description. The problem of course is that it cannot describe the transient dynamics well and is therefore unsuitable when this is what is required. One might try to redefine the Liapunov exponents as local averages as is done in their calculation using the standard algorithm of Bennetin (Pettini 1992). However, in this case, the averages are over temporal states and not trajectories. Such a procedure would not distinguish between non-trivial mixing that leads to evolution and trivial effects such as phase mixing.

Another approach stems from the observation that both the Lagrangian equations of motion and geodesics on the configurational manifold $M$ (the one spanned by the generalized coordinates) of Hamiltonian dynamical systems arise from the same (Maupertuis) variational principle. The Lagrangian description of the system is thus reduced to the study of the properties of this manifold, with complete information on the dynamics (positions *and* velocities) contained in the associated tangent space $TM$ (eg. Arnold 1989). From the Jacoby equation which describes the divergence of infinitely close geodesics on the Riemann space $M$, one can derive the equation for the stability of trajectories to perturbations normal to the geodesic velocity $\parallel \mathbf{u} \parallel$ to be (Anosov 1967)

$$\frac{d^2 \parallel \mathbf{n} \parallel^2}{ds^2} = -k_{\mathbf{u},\mathbf{n}} \parallel \mathbf{n} \parallel^2 + 2 \parallel \nabla_u \mathbf{n} \parallel, \qquad (1)$$

where $k_{\mathbf{u},\mathbf{n}}$ is the two dimensional curvature in a plane defined by $\mathbf{u} \times \mathbf{n}$. If k is negative everywhere on a compact manifold for all $\mathbf{n}$ normal to $\mathbf{u}$ then (1) describes the foliations of expanding and contracting spaces characteristic of Anosov C-systems. Again, one does not expect gravitational systems to possess these characteristics. However there are two properties that make this type of analysis helpful: first we are considering deviations only normal to the motion thus avoiding artifacts arising from differences in phases. Also, the second term in the right hand side of (1) vanishes if one uses Fermi coordinates following the motion (Pesin 1989). Thus estimates of the divergence using this method are *local* and compare divergences of trajectories and not states. Therefore, one may wonder if it is possible to find some averaged criterion that takes advantage of these properties without imposing such strict conditions as required for C-systems. Before investigating such possibilities however I would like to clarify two points that seem to cause misunderstanding. First, it should be clear that *the negativity of the two dimensional curvature on compact manifolds is a sufficient but not necessary condition* for mixing and the approach to equilibrium, these properties belong to all K-systems of which C-systems are but one (extreme) subclass. Second, the negativity of the two dimensional curvature on a subset $m$ of $M$ is a *sufficient condition for chaos on m only if this is the only region visited by the system.*

## 4    Ricci curvature and applications

During its motions a system is continually subjected to perturbations. These will change direction very quickly compared with the evolution time so as to span most directions **n** in the space of directions normal to **u**. To represent this, let $S$ be that space and **Z** be a random vector field uniformly distributed on $S$ with constant magnitude (alternatively one may visualize it as a single vector rotating in $S$). The time evolution of its average magnitude is then given by

$$-\frac{d^2 \int Z d\mathbf{A}}{ds^2} = \int K(\mathbf{A}) Z d\mathbf{A}$$

in Fermi coordinates and where the (Lebesgue) integrals are taken over all directions **A** in $S$. It is well known (eg. Eisenhart 1926) that the quantity

$$r_{\mathbf{u}} = \sum_{\mu=1}^{3N-1} k_{\mathbf{n}_\mu, \mathbf{u}}(s) = R_{ij} \frac{u^i u^j}{\|\mathbf{u}\|^2}$$

(where $R_{ij}$ is the Ricci tensor) does not depend on the particular set of normal directions **n** chosen so that $r_{\mathbf{u}}/(3N-1)$ can be seen as the average value of $k$ over $S$. In addition, $r_{\mathbf{u}}$ does not directly depend on the Riemann tensor, therefore it provides a particularly convenient framework for characterizing instabilities of dynamical systems. Moreover since $Z$ does not depend on **A** we have

$$-\frac{d^2 Z}{ds^2} = r_{\mathbf{u}}/(3N-1) Z. \qquad (2)$$

Negativeness of the Ricci (or mean) curvature $r_{\mathbf{u}}$ will therefore imply that there is average instability under random perturbations. The more negative the curvature the more unstable the system is likely to be (precise definitions are given in Gurzadyan and Kocharyan 1988 and El-Zant 1995). This is a far cry from any rigorous proof of the existence of chaos or the like but is useful for a local quantification of the stability of motion. One can also derive a rough exponentiation timescale for regions of negative curvature

$$\tau_e \sim \left(\frac{3N}{-2\bar{r}_{\mathbf{u}}}\right)^{1/2} \times \frac{1}{\bar{T}} \qquad (3)$$

where the bars represent averages over trajectories and $T$ is the kinetic energy.

Either the full equation (2) or estimates of the type (3) can be used depending on the properties of the system (whether it has mostly negative curvature on its paths). The method can be applied to either individual orbits in fixed potentials, self consistent solutions of the CBE or full $N$-body integrations (perhaps starting from CBE solutions). Various tests have been performed and it was found that for the $N$-Body problem the Ricci curvature is mostly negative when appropriately averaged to get rid of the fluctuations due to close encounters, we describe some of these results in the next section.

## 5 What's wrong with these two body theories?

In El-Zant (1995) the Ricci curvature was calculated for systems of 231 particles started from sheetlike configurations and integrated with very high precision (energy conservation of at least ten digits) for a few dynamical times. The systems studied included collapsing ones which underwent violent relaxation and others which manifested the two types of evolutionary instabilities present in gravitational systems discussed in section 2—namely clustering instabilities and those driving the system towards more symmetric states exhibiting more isotropic velocities. The predictions of the Ricci curvature method were tested against the observed spatial evolution of the different systems both in terms of their relative apparent instability and in terms of absolute timescales. The properties derived from the analysis of the Ricci curvature seemed to agree with the observed behaviour of the systems studied. The effect of softening was also examined. It was found to increase predicted evolutionary timescales considerably.

It is important to check that predictions using the Ricci curvature give the correct description of the approach towards statistical equilibrium when such a state exists, especially since the Ricci curvature is only an averaged quantity. Some investigations were made of softened systems enclosed in "elastic" boxes and which had sufficient energy for stable isothermal spheres to exist. They were started from homogeneous density states in virial equilibrium with (anisotropic) velocities decreasing exponentially with radius. The spheres were divided into ten concentric shells and the trace of the velocity dispersion tensor $\sigma_t$ was calculated in each of them. The relative dispersion of $\sigma_t$

$$\sigma_d = \frac{\sqrt{\frac{1}{N}\Sigma(\sigma_t - \bar{\sigma})^2}}{\bar{\sigma}} \qquad (4)$$

(where the bars denote the average over shells) was calculated as a function of time for systems ranging from $N = 250$ to $N = 2500$. Preliminary results were not found to depend sensitively on $N$, although data for systems with fewer particles was noisier and harder to interpret of course. It was found that these configurations relax towards isothermality in a strikingly short period of a few dynamical times (Fig. 1). Moreover, these state were verified to last for a few hundred dynamical times for systems with $N = 250$ and $N = 500$. Clearly, these results contradict two body relaxation theory. And while there is significant departure from virial equilibrium during the evolution, all evolution cannot be attributed to departure from dynamical equilibrium (violent relaxation) as was verified by considering systems where particles had different masses (Fig. 2). Calculations of the Ricci curvature predict an exponentiation timescale of the order of a dynamical time. This is in good agreement with the results described above for it means that complete relaxation should occur within a few dynamical times.

The above suggests that gravitational systems are close to hyperbolic ones as has been pointed out long ago by Gurzadyan and Saviddy (1986).

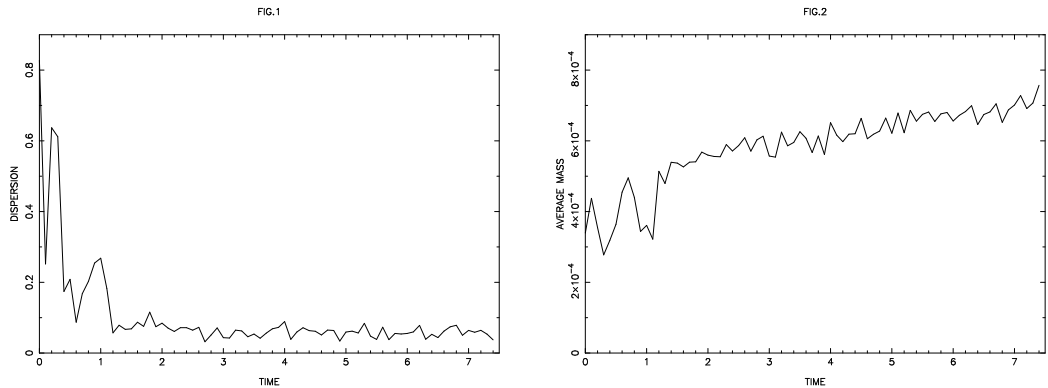

Fig. 1. Relative dispersion of the trace of the velocity dispersion tensor as a function of the crossing time for a system of 2500 equal mass particles as given by eq. (4). Fig. 2. Average mass in innermost cell for system of 2500 particles with salpeter mass function with cutoff mass 10 times greater than the minimum mass and Total mass=1.

## 6  concluding remarks

It is difficult to describe chaos on Non-compact phase spaces—especially ones as complicated as those of large-$N$ systems are likely to be. However, what we hope to obtain using the Ricci curvature method is a *local characterization of the stability of motion*. We have described some tests of the method for some idealized systems, the next step will be to see how it could be applied in realistic situations. Since this is obviously not a trivial problem one may start by examining the stability of $N$-body realizations of systems starting near steady state solutions of the CBE—that is mapping out this region of the phase-space. This type of analysis should be a powerful aid in the interpretation of the results of short time integrations of such systems which might help in the understanding of their evolutionary behaviour.